\documentclass{appolb}
\usepackage[sort&compress, numbers, square, comma]{natbib}

\usepackage{amsmath,amssymb}
\usepackage{color,url}
\usepackage{listings}
\usepackage{slashed}
\usepackage[pdftex]{graphicx}
\usepackage{epstopdf}
\usepackage{epsfig}
\usepackage{grffile}
\usepackage{relsize}
\usepackage{float}
\graphicspath{{./img/}}
\usepackage{sidecap}
\usepackage{soul}

\usepackage{CJKutf8}

\usepackage[mathscr,scaled=1.15]{urwchancal}
\DeclareFontFamily{OT1}{pzc}{}
\DeclareFontShape{OT1}{pzc}{m}{it}%
{<-> s * [1.15] pzcmi7t}{}
\DeclareMathAlphabet{\mathpzc}{OT1}{pzc}{m}{it}

\definecolor{purple}{rgb}{0.5,0,0.5}
\definecolor{blue}{rgb}{0.0,0,0.9}
\definecolor{prdblue}{rgb}{0.133,0.118,0.498}
\usepackage[colorlinks=true, pdfstartview=FitV, linkcolor=prdblue, citecolor= prdblue, urlcolor=prdblue]{hyperref}

\RequirePackage{comment}

\begin{document}
\begin{CJK*}{UTF8}{gbsn}

\title{\vspace*{-2ex}
\hspace*{\fill}{\normalsize{\sf\emph{Preprint no}.\ NJU-INP 108-25}}\\
$\rule{0ex}{1ex}$\\
Insights into Nucleon Resonances \\ via Continuum Schwinger Function Methods
}

\author{
    Peng Cheng (程鹏),%
    $^{\href{https://orcid.org/0000-0002-6410-9465}{\textcolor[rgb]{0.00,1.00,0.00}{\sf ID}}\,1\!\!}$
\thanks{pcheng@mail.ahnu.edu.cn}
Langtian Liu (刘浪天),%
    $^{\href{https://orcid.org/0000-0002-4283-0315}{\textcolor[rgb]{0.00,1.00,0.00}{\sf ID}}\,2\!\!}$
\thanks{liult@stu.edu.cn}
\\
Ya Lu (陆亚),%
       $\,^{\href{https://orcid.org/0000-0002-0262-1287}{\textcolor[rgb]{0.00,1.00,0.00}{\sf ID}}\,3\!\!}$
\thanks{luya@njtech.edu.cn}
Craig D.\ Roberts%
       $^{\href{https://orcid.org/0000-0002-2937-1361}{\textcolor[rgb]{0.00,1.00,0.00}{\sf ID}}\,4,5\!\!}$
\thanks{cdroberts@nju.edu.cn}
\address{$^1$Department of Physics, \href{https://ror.org/05fsfvw79}{Anhui Normal University}, Wuhu, Anhui 24100, China\\
$^2$Physics Department, College of Science, \href{https://ror.org/01a099706}{Shantou University}\\
Shantou, Guangdong 515063, China\\
$^3$Department of Physics, \href{https://ror.org/03sd35x91}{Nanjing Tech University}, Nanjing 211816, China \\
%
$^4$School of Physics, \href{https://ror.org/01rxvg760}{Nanjing University}, Nanjing, Jiangsu 210093, China\\
$^5$Institute for Nonperturbative Physics, \href{https://ror.org/01rxvg760}{Nanjing University}, \\ Nanjing, Jiangsu 210093, China
\date{2025 December 10\\Contribution to a special issue that celebrates the 90$^{\it \, th}$ birthday of L.\ David Roper}
\\
}}

\maketitle

\begin{abstract}
The first baryon resonance was discovered in the early 1950s.  The Roper resonance joined the collection ten  years later.  Today, many baryon resonances are known and more are being discovered.  As baryons, these states are the most fundamental three-body systems in Nature.  They must all be understood, not just the isolated ground state nucleon.  This contribution sketches applications of continuum Schwinger function methods to the baryon resonance problem.  Whilst spectroscopy is of value, particular emphasis is placed on resonance electroproduction because transition form factors extracted from electroproduction data provide a keen tool for revealing resonance structure.
\end{abstract}
\end{CJK*}

\section{Introduction}
Around forty years ago, it was common for practitioners to judge that (constituent) quark potential models were providing a realistic picture of the baryon spectrum; indeed, that they were a ``\emph{phenomenal phenomenological success}'' \cite{Hey:1982aj}.
Moreover, there was a perception \cite{Hey:1982aj} that: ``\emph{Although there may still be some weakly coupled resonances lurking around the noise level or background of partial-wave analyses, it seems clear that the major features of the spectrum are known. It is not at all clear that we will ever have much more than, at best, a rudimentary outline of charmed or bottom baryon spectroscopy and it is probable that we have now identified over 90\% of the resonant states that we shall ever disentangle from the experimental data.  Indeed, given present experimental trends, it seems probable also that little more, if any, new experimental data relevant to the baryon spectrum will be forthcoming}.''
Now we know that such conclusions were premature.  This is made clear by the continuing discovery of new baryon resonances \cite{Burkert:2019kxy, Mokeev:2020hhu} and states considered to be tetraquark or pentaquark systems that contain heavy valence quarks \cite{Liu:2019zoy, Chen:2022asf}.
Today, therefore, the study of the nucleon and its resonances is an active area of research in both experiment and theory.

That is rightly the case because baryons and their resonances play a central role in the existence of our universe and ourselves; and therefore \cite{Isgur:2000ad}: ``\emph{\ldots\ they must be at the center of any discussion of why the world we actually experience has the character it does.}''  Baryons are the most fundamental three-body systems in Nature.  They are supposed to be described by quantum chromodynamics (QCD) \cite{Fritzsch:1973pi}.  Yet, fifty years after the introduction of QCD, it is impossible to claim that we even understand proton and neutron structure; and the number and structure of nucleon resonances remains a topic that attracts much debate.
(The picture for pions and kaons, Nature's most fundamental Nambu-Goldstone bosons, is more parlous still \cite{Ding:2022ows, Roberts:2021nhw}.  This is recognised and the goal of revealing pion and kaon structure is becoming an intense focus of new experiments \cite{Roberts:2021nhw, Chen:2020ijn, Arrington:2021biu, Quintans:2022utc}.)

Considering baryon resonances constituted from some three-body combination of $u$, $d$, $s$ valence quarks, the quark model spectroscopic labelling convention is still popular, namely, cataloguing hadrons as $n\,^{2s+1}\!\ell_J$ systems, where $n$, $s$, $\ell$ are radial, spin, and orbital angular momentum quantum numbers, with $\ell+s=J$ and $J$ being the total angular momentum.
In fact, it is sometimes/often reported that \cite[Ch.~15]{ParticleDataGroup:2024cfk}:
(\emph{a}) The spectrum of baryons and mesons exhibits a high degree of regularity.
(\emph{b}) The organisational principle which best categorizes this regularity is encoded in the quark model.
(\emph{c}) All descriptions of strongly interacting states use the language of the quark model.

Point (\emph{a}) is debatable.  Many baryons are known, but the perceived degree of regularity is practitioner dependent, \emph{e.g}., the existence and/or character of Regge trajectories is certain for some practitioners but tenuous for others \cite{Tang:2000tb, Masjuan:2012gc}.
In this connection, the perspective in Ref.\,\cite{Veltmann:2003} is valuable:
`\emph{`In time the Regge trajectories thus became the cradle of string theory.  Nowadays the Regge trajectories have largely disappeared, not in the least because these higher spin bound states are hard to find experimentally.  At the peak of the Regge fashion (around 1970) theoretical physics produced many papers containing families of Regge trajectories, with the various (hypothetically straight) lines based on one or two points only}!''  These observations remain pertinent to  contemporary models that support Regge trajectories.

Given that (\emph{a}) is questionable, then (\emph{b}) is merely the perspective of an embedded community.

Point (\emph{c}) is false: Poincar\'e-covariant treatments of the baryon spectrum need not make any reference to quark model language.
Indeed, since $u$, $d$, $s$ valence quarks are light, even the concept of a QCD-connected ``potential'' with any connection to Schr\"odinger equation quantum mechanics is unsound, so no such potential model labelling scheme is admitted.

To these remarks, one should add that Poincar\'e covariance itself, required for any QCD-connectable treatment of bound states seeded by light valence quarks, entails that every hadron contains orbital angular momentum, \emph{e.g}.,
even the ground-state pion contains two {\sf S}-wave and two {\sf P}-wave components
\cite{Bhagwat:2006xi, Hilger:2015ora}.
Thus, Nature's pion cannot be a $1\,^1\!S_0$ state in any sense.
Moreover, no system is simply a radial excitation of another: in a Poincar\'e-covariant wave function, there are simply too many degrees of freedom for that to be the case.
On top of these things, whilst $J$ is Poincar\'e invariant -- so, truly an observable -- any separation of $J$ into orbital angular momentum plus spin, $J=L+S$, is frame dependent, \emph{i.e}., subjective.
Continuing along this line, it should be borne in mind that, in quantum field theory, orbital angular momentum and parity are unconnected.
This is plain because parity is a Poincar\'e invariant quantum number, whereas $L$ is not; and no observable can properly be defined by a subjective quantity.
Consequently, negative parity states cannot simply be orbital angular momentum excitations of positive parity ground states.
These features of the Poincar\'e covariant treatment of baryons are detailed elsewhere; see, \emph{e.g}., Refs.\,\cite{Eichmann:2016yit, Chen:2017pse, Qin:2018dqp, Qin:2019hgk, Barabanov:2020jvn, Liu:2022ndb, Liu:2022nku, Eichmann:2022zxn}, and some will be reiterated below.

\section{Roper Resonance}
\label{SecRoper}
Many (most) readers of this contribution will know that the Roper resonance was discovered in 1963 \cite{Roper:1964zza, BAREYRE1964137, AUVIL196476, PhysRevLett.13.555, PhysRev.138.B190}; and a large subset of that group will acknowledge that the Roper's characteristics have been the source of great puzzlement since that time.

Here, it is therefore appropriate to state the simplest of these characteristics; namely, the Roper is a spin-half, positive-parity resonance, $J^P = 1/2^+$, with pole mass $\approx 1.37\,$GeV and width $\approx 0.18\,$GeV \cite{Olive:2016xmw}.  In the spectrum of nucleon-like states, \emph{i.e}., baryons with isospin
%
$I=1/2$, the Roper resonance lies about $0.4\,$GeV above the ground-state nucleon and $0.15\,$GeV below the first spin-half negative-parity state, $J^P =1/2^-$, which has roughly the same width.  Today, the levels in this spectrum are labelled thus:
$N({\rm mass})\,J^P$;
and hence the ground-state nucleon is denoted $N(940)\,1/2^+$, the Roper resonance as $N(1440)\,1/2^+$, and the negative-parity state described above is $N(1535)\,1/2^-$.

The first Poincar\'e-covariant treatment of the Roper resonance was completed ten years ago \cite{Segovia:2015hra}.
It employed a continuum approach to the three valence-quark bound-state problem in relativistic quantum field theory to predict a range of properties of the proton's radial excitation and thereby unify them with those of numerous other hadrons.
The analysis indicated that the Roper resonance should be identified with the nucleon's first radial excitation.
Moreover that, structurally, the resonance is built from a core of three dressed quarks, which expresses its valence-quark content and whose charge radius is 80\% larger than the proton analogue; and that quark core is complemented by a meson cloud, which reduces the observed Roper mass by roughly 20\% \cite{Suzuki:2009nj}.
The meson cloud materially affects long-wavelength characteristics of the Roper electroproduction amplitudes but the quark core is revealed to probes with $Q^2 \gtrsim 2 m_N^2$, where $m_N$ is the nucleon mass.

It was only possible for the Ref.\,\cite{Segovia:2015hra} study to reach this conclusion following the accumulation and analysis of high-precision $N(940)\to N(1440)$ electroproduction data by the CLAS Collaboration at Jefferson Laboratory (JLab).
As explained in Refs.\,\cite{Burkert:2019bhp, Mokeev:2025hhe}, spectroscopy alone cannot reveal structural features of any given set of bound states: for instance, no one would claim to know and understand proton structure merely because they had arrived at a value of its mass.
Notwithstanding these things, some groups, which focus on spectroscopy, still suggest that an alternative (meson + nucleon molecule) explanation of Roper structure is viable, without having tested their picture against electroproduction data.
Herein, therefore, it is natural to recapitulate and expand on some of the key points developed in Ref.\,\cite{Segovia:2015hra}.

\begin{figure}[t]
\centerline{%
\includegraphics[clip, width=0.70\textwidth]{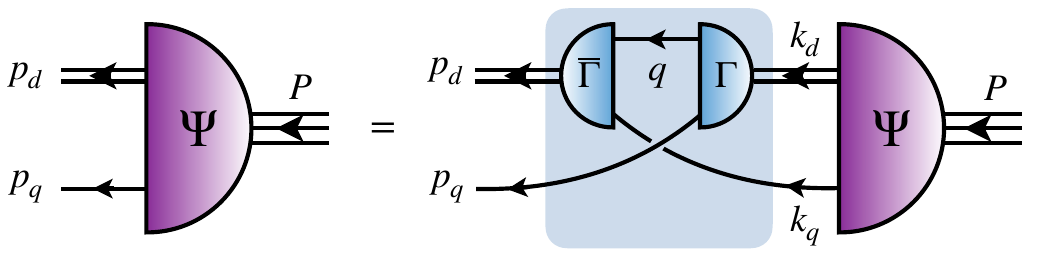}}
\caption{\label{F1}
Linear, homogeneous integral equation for $\Psi$, the Poincar\'e-covariant matrix-valued function (Faddeev amplitude) for a baryon with total momentum $P=p_q+p_d=k_q+k_d$ constituted from three valence quarks, two of which are paired in a fully-interacting nonpointlike diquark correlation.
$\Psi$ expresses the relative momentum correlation between the dressed-quarks and -diquarks.
Legend. \emph{Shaded box} -- Faddeev kernel, which explicitly shows the quark exchange binding mechanism;
\emph{single line} -- dressed-quark propagator;
$\Gamma$ -- diquark correlation amplitude;
and \emph{double line} -- diquark propagator..
}
\end{figure}

In QCD, the properties of the proton (nucleon) and its resonances can be studied in any approach that provides access to the three-quark six-point Schwinger function \cite{SW80, GJ81}.
In this connection, continuum Schwinger function methods (CSMs) provide a widely used and insightful calculational scheme \cite{Eichmann:2016yit, Binosi:2022djx, Ding:2022ows, Ferreira:2023fva}.
Many such studies use a quark + dynamical diquark -- $q(qq)$ -- picture of baryons because it vastly simplifies the problem \cite{Barabanov:2020jvn}.  This was the approach employed in Ref.\,\cite{Segovia:2015hra}.

\begin{figure}[t]
\hspace*{-1ex}\begin{tabular}{lcl}
\large{\textsf{A}} & & \large{\textsf{B}}\\[-0.7ex]
%
\includegraphics[clip, width=0.45\textwidth]{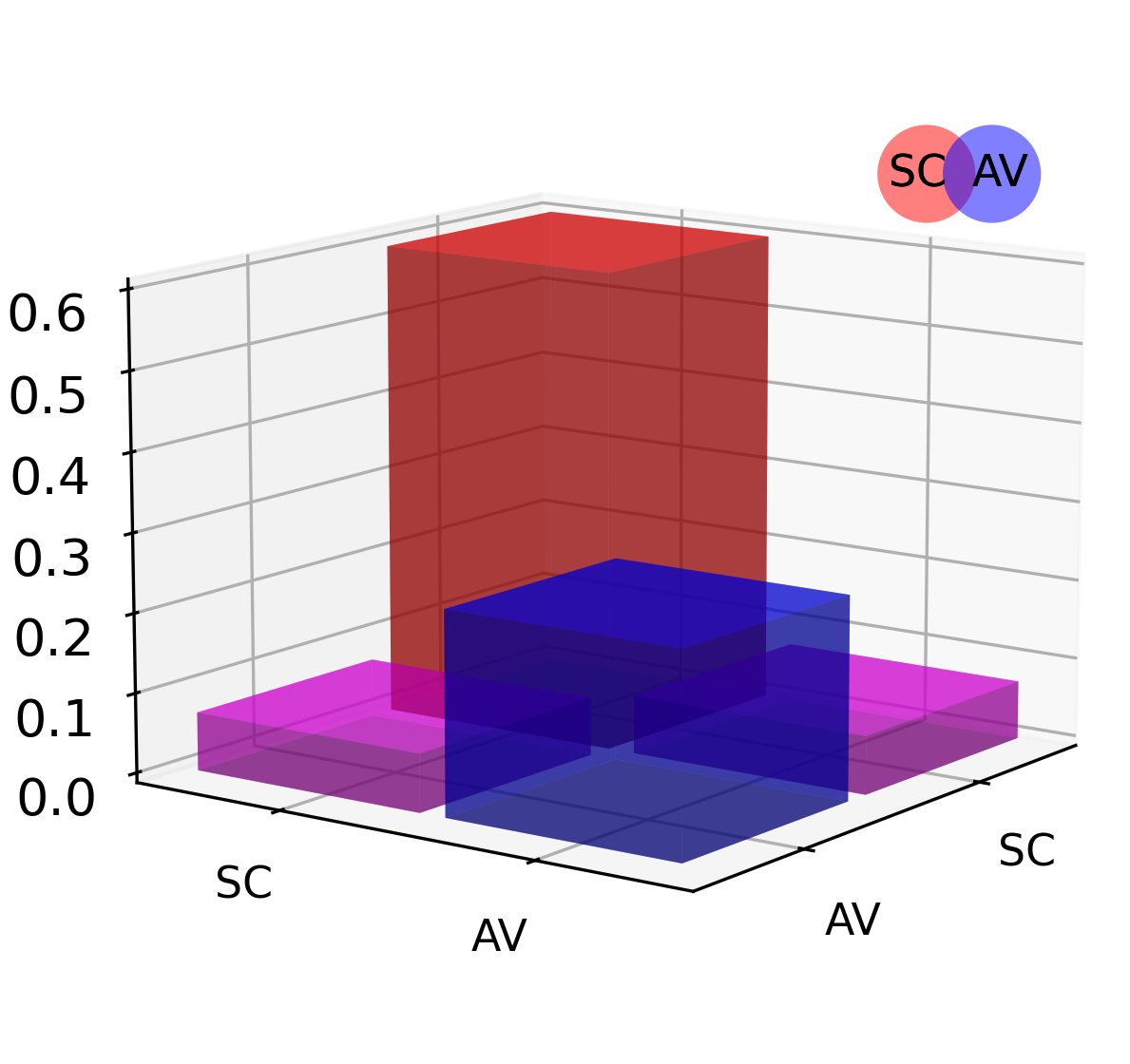} & \hspace*{1em} &
\includegraphics[clip, width=0.475\textwidth]{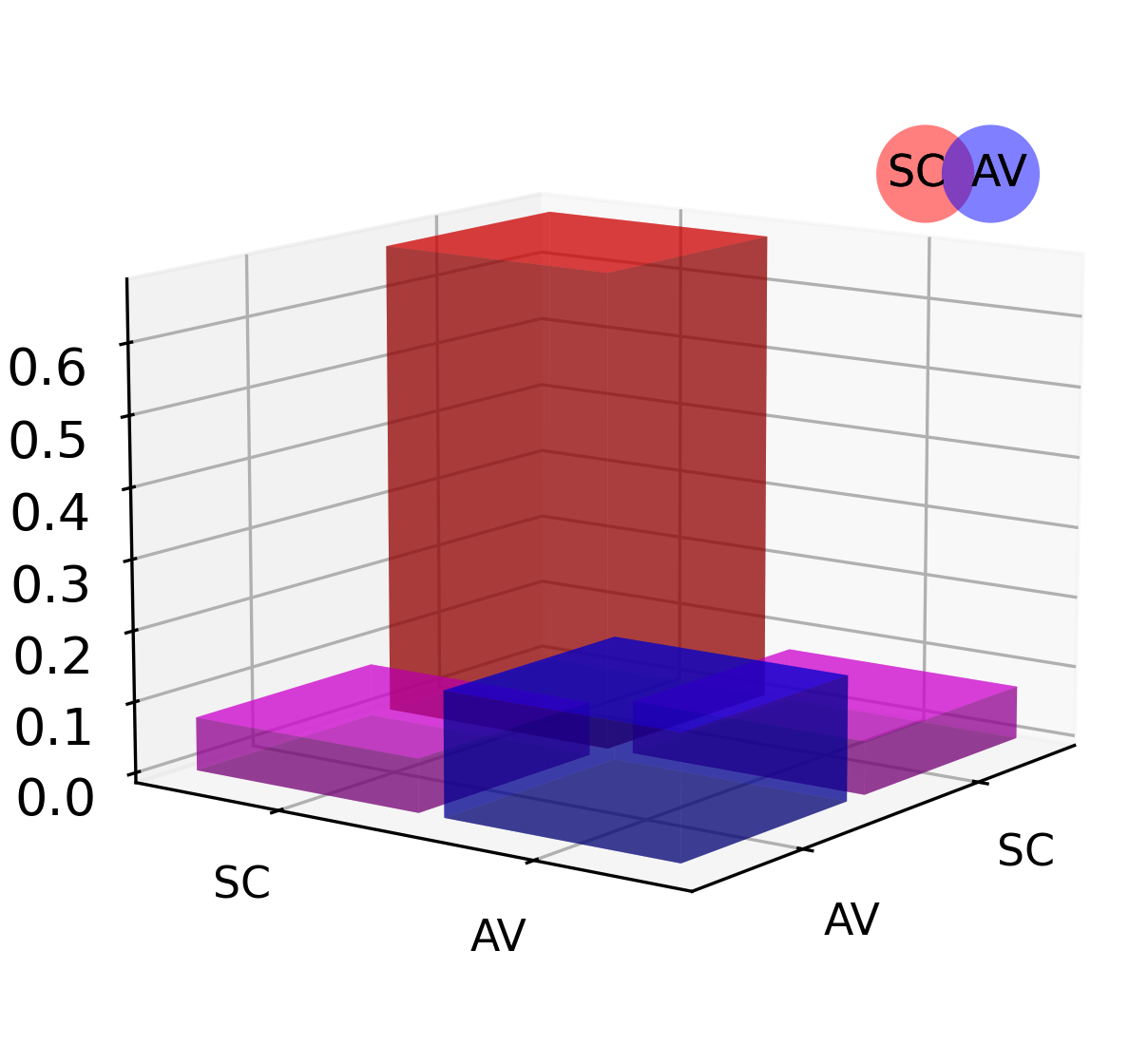}
\end{tabular}
\caption{\label{F2}
{\sf Panel A}.
Proton.  Diquark component breakdown of the canonical normalisation of the proton's Poincar\'e-covariant nucleon Faddeev wave function.  The $[ud]_{0^+}$ isoscalar-scalar diquark (SC) is dominant, but material contributions also owe to the $\{uu\}_{1^+}$, $\{ud\}_{1^+}$ isovector-axialvector correlations (AV).
SC$\,\otimes\,$SC -- 60\%;
SC$\,\otimes\,$AV -- 15\%;
AV$\,\otimes\,$AV -- 25\%.
{\sf Panel B}. Analogous image for the first $1/2^+$ excitation of the ground-state nucleon.
SC$\,\otimes\,$SC -- 67\%;
SC$\,\otimes\,$AV -- 15\%;
AV$\,\otimes\,$AV -- 18\%.
%
%
}
\end{figure}

The $q(qq)$ picture of baryon structure was introduced in Refs.\,\cite{Cahill:1988dx, Reinhardt:1989rw, Efimov:1990uz}, which exploited the pairing capacity of fermions as a means by which to simplify the three valence-quark bound state problem.
The analyses therein lead to a Faddeev equation that describes dressed quarks and fully-interacting diquark correlations built therefrom, which bind together into a baryon, at least in part, because of the continual exchange of roles between the spectator and diquark-participant quarks.
The associated bound state equation is sketched in Fig.\,\ref{F1}, which explicitly shows the action of diquark breakup and reformation in the kernel.
First solutions of the $q(qq)$ Faddeev equation were described in Ref.\,\cite{Burden:1988dt}.
Since then, the approach has evol\-ved into a sophisticated tool that has been used to predict many baryon observables; see, \emph{e.g}., Refs.\,\cite{Chen:2023zhh, Yu:2025fer, Xu:2025ups}.
The most refined version of the approach can be found in Ref.\,\cite{Cheng:2025yij}.

Since calculational details are recorded in Refs.\,\cite{Segovia:2015hra, Cheng:2025yij}, we turn immediately here to a discussion of the nucleon and Roper Poincar\'e-covariant wave functions.
The first thing to note is the diquark content.
The analyses in Refs.\,\cite{Segovia:2015hra, Chen:2019fzn} reveal that isoscalar-scalar and isovector-axialvector diquarks are necessary to complete a stable (mass-converged) solution of the Faddeev equation.
Regarding their relative contributions to the canonical normalisations, both the scalar and axialvector are significant in the nucleon and Roper wave functions.
In fact, the $(I=0,0^+)$ diquark contributes approximately 60-70\% to the normalisation and the $(1,1^+)$ diquark, both directly and through constructive interference with the $(0,0^+)$ correlation, provides the remaining 30-40\%.  For the proton, this is illustrated in Fig.\,\ref{F2}\,A and for the Roper, in Fig.\,\ref{F2}\,B:
plainly, their diquark substructure is very similar.
(The canonical normalisation constant is fixed by the requirement that the $Q^2=0$ value of the charge form factors associated with electrically charged members of a given hadron multiplet deliver the observed charge.  For the proton and charged Roper, this condition is expressed in $F_1(Q^2=0)=1$, where $F_1$ is the Dirac form factor of the proton or charged-Roper.)

\begin{figure}[t]
\begin{minipage}[t]{\textwidth}
\leftline{%
\includegraphics[clip, width=0.5\textwidth]{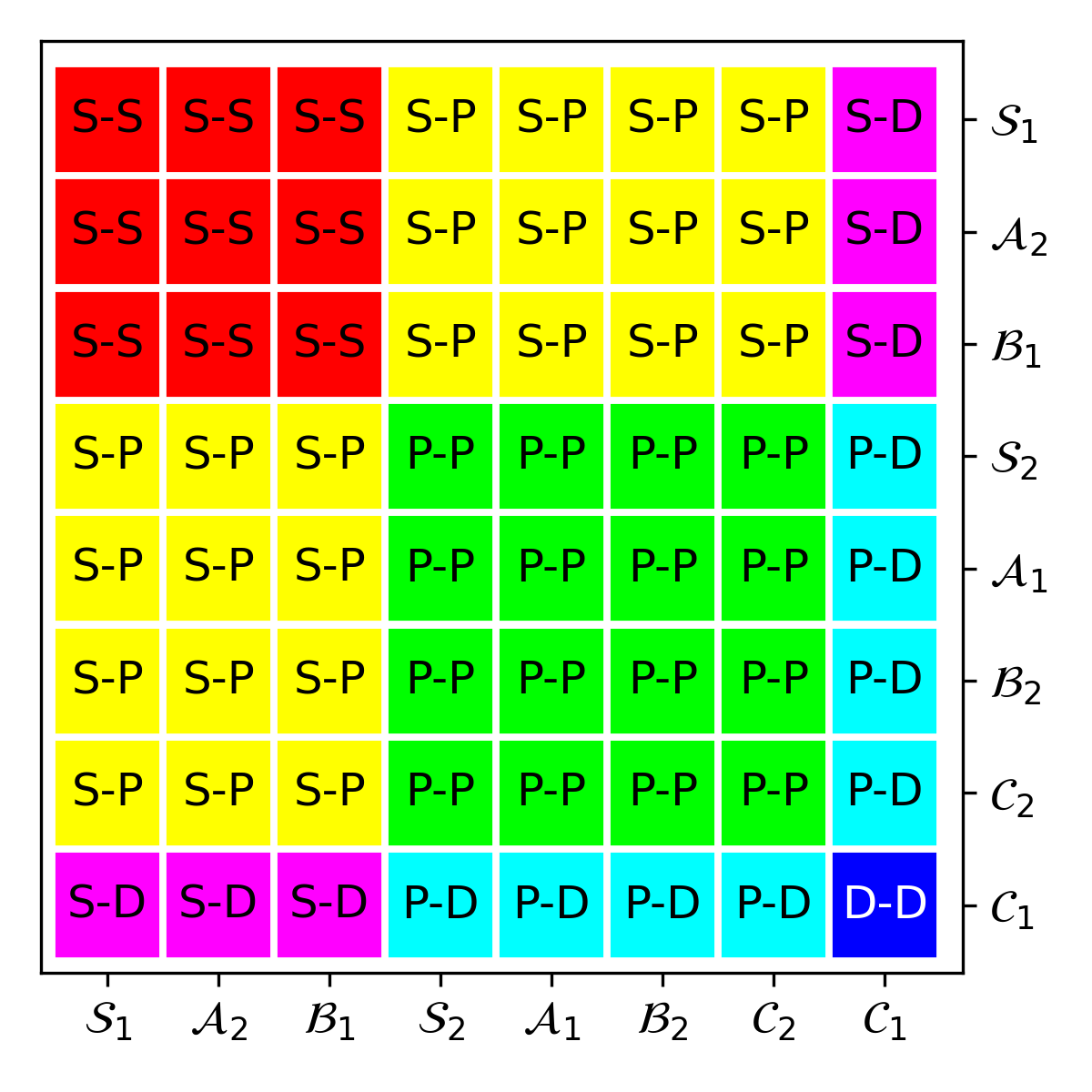}}
\end{minipage}\vspace*{-17em}
\rightline{
\begin{minipage}[t]{0.45\textwidth}
\caption{\label{F3}
Legend for interpretation of $J=1/2$ baryon rest-frame quark + diquark angular momentum decompositions, which also identifies interference between the distinct orbital angular momentum basis components.
The axes labels refer to distinct components of the $q(qq)$ wave function, which are explicitly detailed in
Ref.\,\cite[Eq.\,(11)]{Chen:2019fzn}.
}\vspace*{2em}
\end{minipage}
}
\end{figure}

\begin{figure}[t]
\hspace*{-1ex}\begin{tabular}{lcl}
\large{\textsf{A}} & & \large{\textsf{B}}\\[-0.7ex]
%
\includegraphics[clip, width=0.45\textwidth]{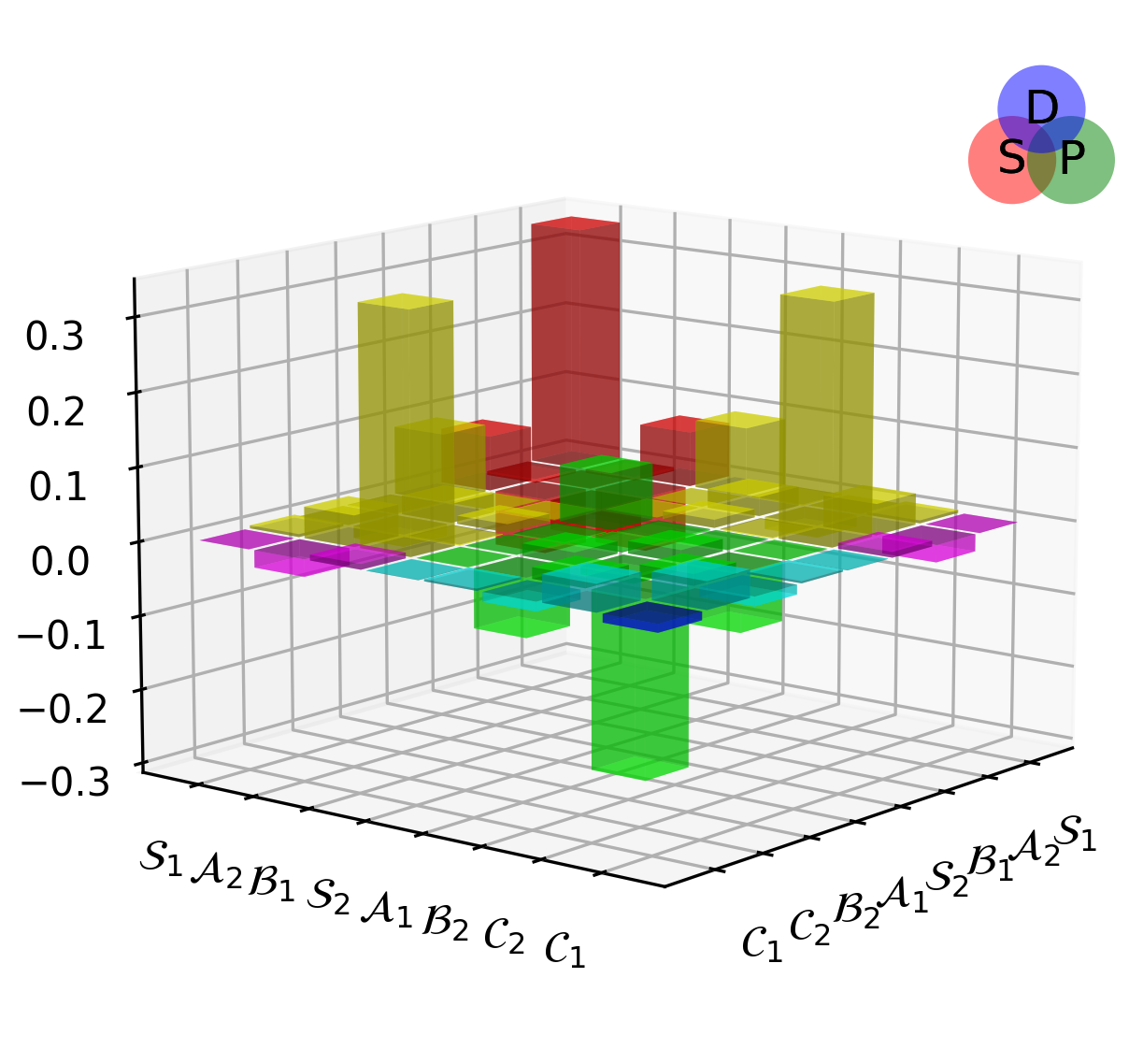} & \hspace*{1em} &
\includegraphics[clip, width=0.45\textwidth]{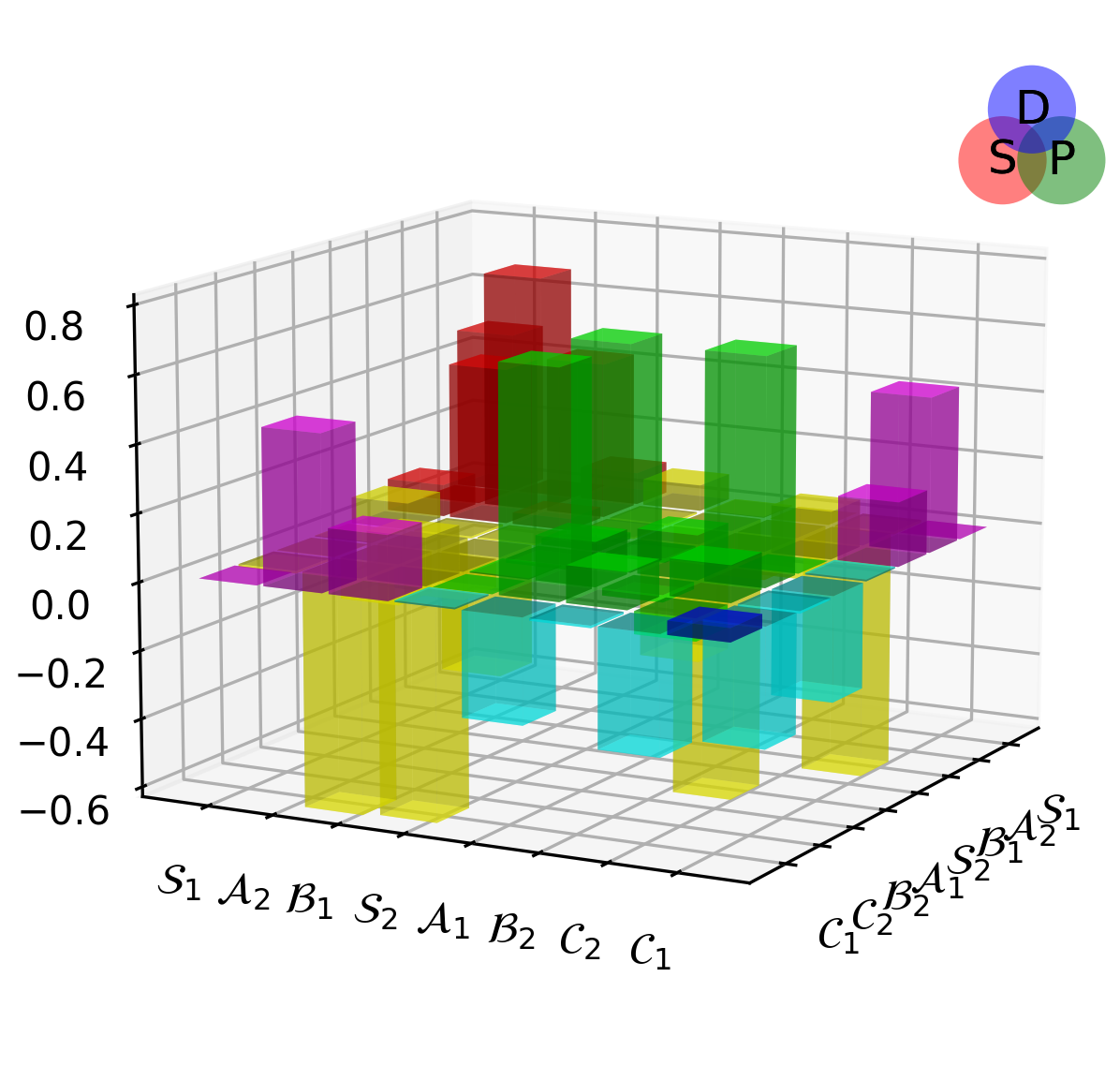}
\end{tabular}
\hspace*{-1ex}\begin{tabular}{lcl}
\large{\textsf{C}} & & \large{\textsf{D}}\\[-0.7ex]
%
\includegraphics[clip, width=0.45\textwidth]{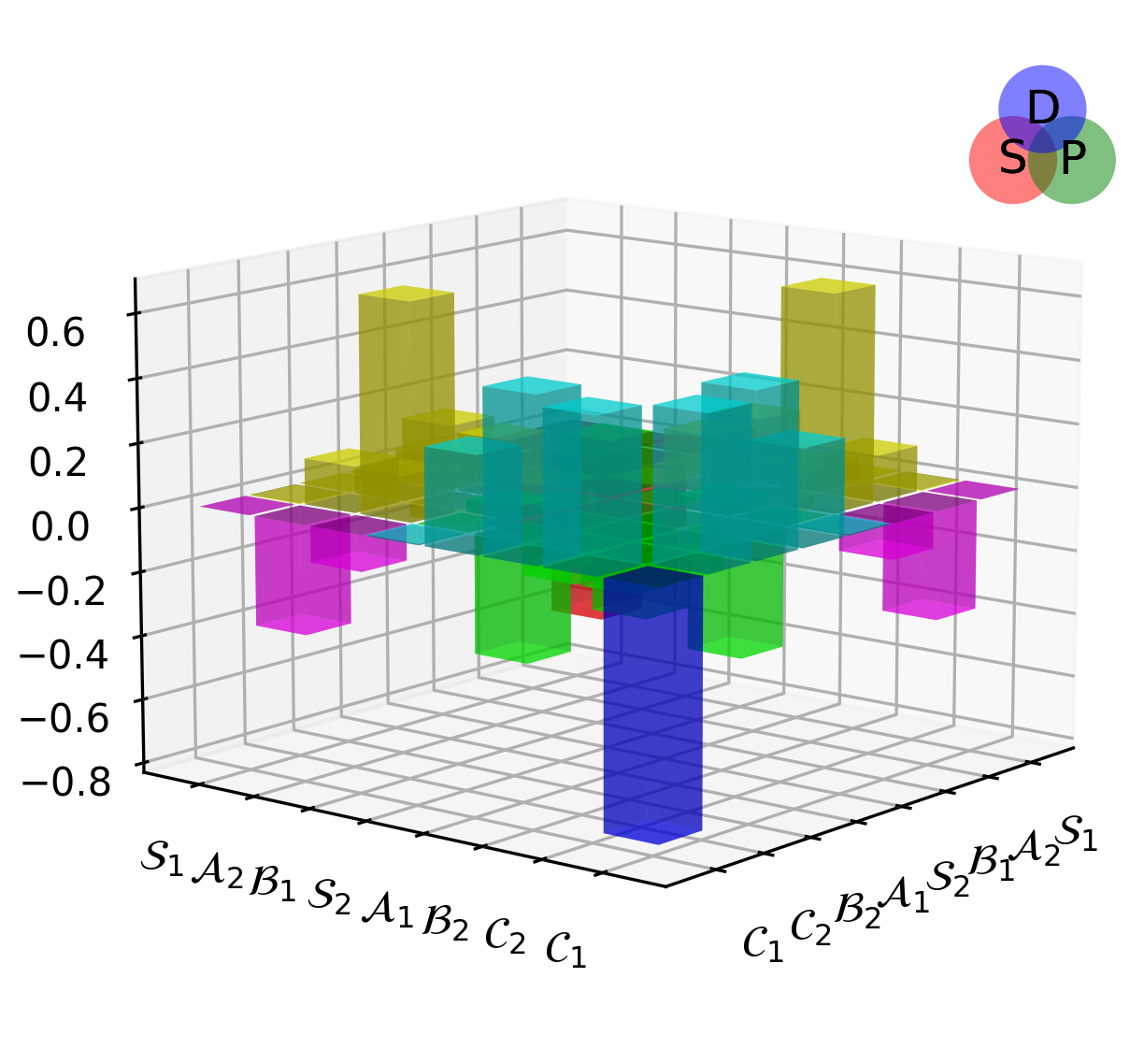} & \hspace*{1em} &
\includegraphics[clip, width=0.45\textwidth]{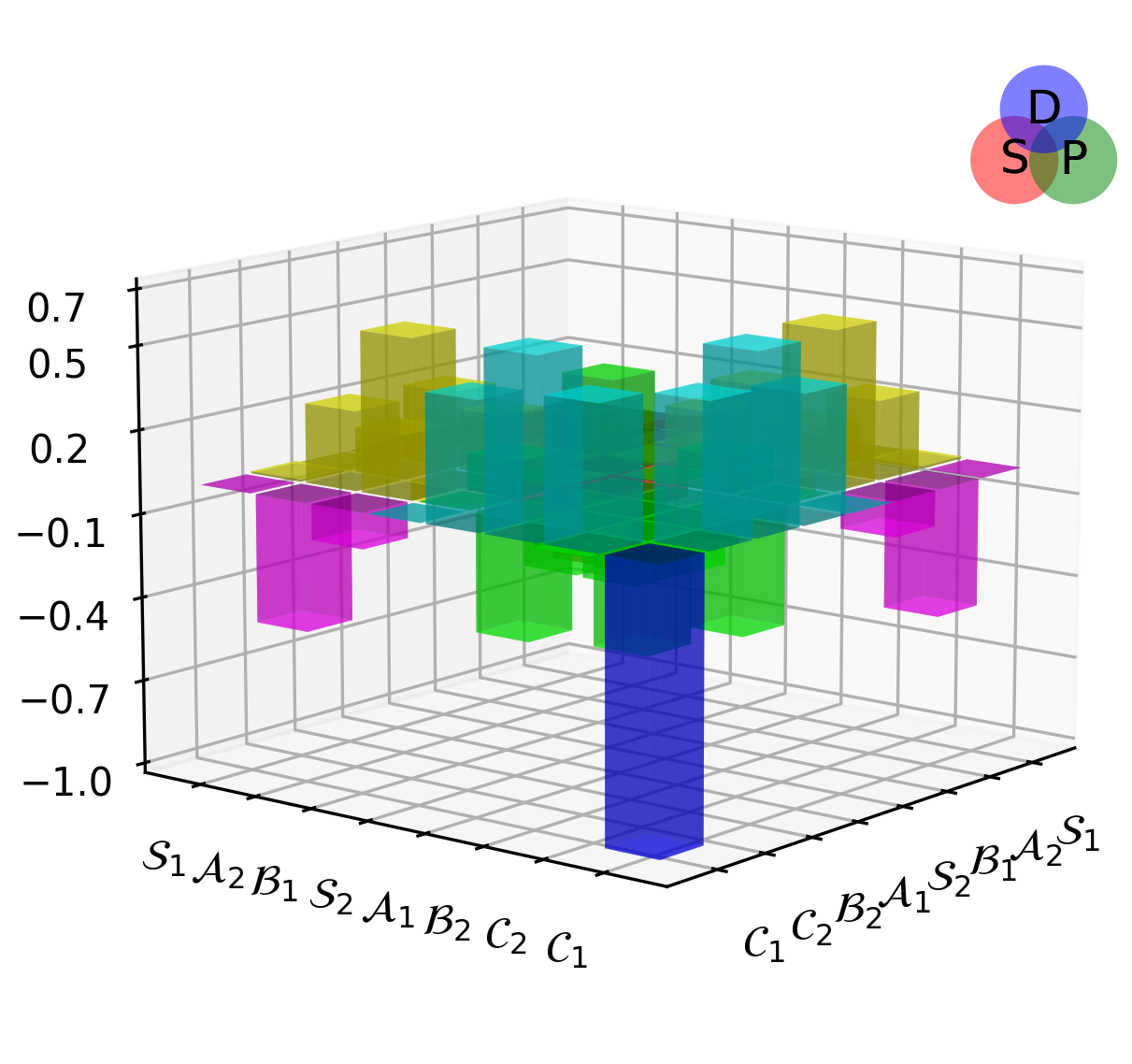}
\end{tabular}
\caption{\label{F4}
Contributions of the various quark + diquark orbital angular momentum components to the canonical normalisation of the Poincar\'e-covariant wave function of a $J=1/2$ baryon after rest-frame projection: there are both positive (above plane) and negative (below plane) contributions to the overall positive normalisation.
{\sf Panel A}.  Proton.
{\sf Panel B}.  First $J=1/2^+$ excitation of the proton, identified as the Roper resonance.
{\sf Panel C}.  First $J=1/2^-$ excitation of the proton, identified as the $N(1535)1/2^-$.
{\sf Panel D}.  Second $J=1/2^-$ excitation of the proton, identified as the $N(1650)1/2^-$.
%
%
}
\end{figure}

Of particular additional interest are pictures of the rest-frame angular momentum structure of the nucleon and Roper resonance.  Referring to the legend in Fig.\,\ref{F3}, these decompositions are drawn in Fig.\,\ref{F4}.
Panel\,\ref{F4}\,A displays the proton case.  Plainly, the proton wave function has significant ${\mathsf S}$-wave components; yet it also contains material ${\mathsf P}$-wave structures and the canonical normalisation receives measurable ${\mathsf S}\otimes {\mathsf P}$-wave interference contributions.
The verity of this picture of the proton wave function is confirmed, \emph{e.g}.,
by its successful explanation of 
nucleon elastic electromagnetic and gravitational form factors \cite{Cheng:2025yij, Yao:2024uej, Yao:2024ixu}
and a large array of unpolarised and polarised proton parton distribution functions \cite{Lu:2022cjx, Cheng:2023kmt, Yu:2024ovn}.

Turning now to Fig\,\ref{F4}\,B, one sees the rest-frame quark + diquark angular momentum projection of the Poincar\'e-covariant wave function of the proton's first $1/2^+$ excitation.
The wave function of this system has properties that justify its interpretation as the proton's first radial excitation.  For instance, the zeroth Chebyshev moment of each component possesses a single zero \cite{Segovia:2015hra}.
So, as explained in connection with meson radial excitations \cite{Holl:2004fr, Li:2016dzv}, this state is properly identified in quantum field theory with the radial excitation of the $N(940)$.  (Transition form factors will soon be discussed.)

On the other hand, in the Poincar\'e-covariant $q(qq)$ approach, the first $1/2^+$ excitation is obviously far more than simply a radial excitation.
Indeed, its $q(qq)$ angular momentum structure is completely different from that of the proton:
the ${\mathsf S}$- and ${\mathsf P}$-wave components are enhanced in magnitude, with ${\mathsf P}$-waves flipped in sign;
the ${\mathsf S}\otimes {\mathsf P}$-wave interference terms are, in this state, strongly destructive
as are ${\mathsf P}\otimes {\mathsf D}$-wave terms;
and there are strong constructive ${\mathsf S}\otimes {\mathsf D}$-wave contributions.
Notwithstanding these significant differences, identification of this system with the quark core of the Roper resonance is justified by the fact that the wave function which yields Fig\,\ref{F4}\,B also delivers agreement with the $\gamma^\ast N(940) \to N(1440)$ transition form factors on $Q^2 \gtrsim 2m_N^2$ \cite{Segovia:2015hra, Burkert:2019bhp}.
Notably, the predictions in Ref.\,\cite{Segovia:2015hra} have been updated.
The new results will be reported elsewhere.  Here it is sufficient to record that, without reference to Roper electroproduction data in its formulation, the framework in Ref.\,\cite{Cheng:2025yij} delivers even better results than those in the first study \cite{Segovia:2015hra}.

It is worth augmenting our analysis of the upper images in Fig.\,\ref{F4} with a discussion of the lower panels, which depict rest-frame angular momentum projections of the Poincar\'e-covariant wave functions of the $N(1535)1/2^-$ and $N(1650)1/2^-$.
In quark potential models, these states are both orbital angular momentum excitations of the ground state nucleon: $N(1535)1/2^-$ -- $(L=1, S=1/2)$; $N(1650)1/2^-$ -- $(L=1, S=3/2)$.
Evidently, considering Figs.\,\ref{F4}\,C, D, the quantum field theory pictures are more complex.
In both cases, to somewhat differing degrees:
(\emph{a}) $\mathsf P$-wave components are significant, but they make a negative contribution to the normalisation;
(\emph{b}) the largest constructive contributions owe to
${\mathsf S}\otimes {\mathsf P}$-wave and ${\mathsf P}\otimes {\mathsf D}$-wave interference;
and
(\emph{c}) ${\mathsf S}\otimes {\mathsf D}$ and ${\mathsf D}\otimes {\mathsf D}$ terms are negative.
The $N(1535)1/2^-$ wave function has been validated by comparisons of data with predictions for the $\gamma^\ast N(940) \to N(1535)1/2$ transition form factors, which will soon be released.
(An existing $q(qq)$ analysis, based on a symmetry-preserving, Poincar\'e-covariant treatment of a vector$\,\otimes\,$vector contact interaction, sets a baseline for such studies \cite{Raya:2021pyr}.)
Analogous tests are currently underway in connection with the $N(1650)1/2^-$.

It should also be remarked that, in quantum field theory, the $N(1535)1/2^-$ and $N(1650)1/2^-$ are properly viewed as parity partners of the nucleon.
Thus, with respect to the $N(940)$, any and all differences in mass and structure owe to dynamical chiral symmetry breaking -- see, \emph{e.g}., Refs.\,\cite{Wang:2013wk, Aarts:2015mma, Aarts:2017rrl, Chen:2017pse}, which is a corollary of emergent hadron mass (EHM) \cite{Roberts:2016vyn, Krein:2020yor, Roberts:2021nhw, Binosi:2022djx, Ding:2022ows, Ferreira:2023fva, Salme:2022eoy, Achenbach:2025kfx}.
The similarities between the rest-frame $q(qq)$ angular momentum projections of the Poincar\'e-covariant wave functions of these states supports this perspective.
Any claim, today, that parity-doubling (degeneracy of parity partners) is seen in the baryon spectrum is false: instead, it merely expresses a misapprehension of the true character of parity partners.
Absent EHM, on the other hand, the $N(1535)1/2^-$ and $N(1650)1/2^-$ resonances would both be degenerate with the ground-state nucleon and have identical wave functions, \emph{i.e}., these three states would be indistinguishable.  The same would then, and only then, be true of all real parity partners.
Equivalent statements hold for mesons \cite{Weinberg:1967kj, Chang:2011ei}.

The images in Fig.\,\ref{F4} highlight that any approach to calculating properties of the nucleon and its excitations which fails to express the complex character of their wave functions, will very likely deliver an erroneous structural picture of these states.
Notably, the much debated $N(1440)$ is both a radial and orbital angular momentum excitation of the $N(940)$, and the same is true of the other states.

Focusing on the Roper, now that electroexcitation data exist \cite{Burkert:2019bhp}, proponents of alternative interpretations of the Roper resonance can subject their pictures to stringent validation tests.  Absent such tests, it is rational to question any claims that the Roper is unrelated to the proton's first radial excitation.  The future will see robust calculations based on the three valence body Faddeev equation ($3$-body) without reference to diquark correlations \cite{Eichmann:2011vu, Qin:2019hgk, Yao:2024uej}.  To date, wherever comparisons have been made, $3$-body and $q(qq)$ results are largely in agreement \cite{Eichmann:2016yit, Cheng:2025yij, Xu:2025ups}.

\section{$\Delta(1600)3/2^+$}
The above discussion has highlighted that the QCD structure of hadrons is far richer than can be produced by quark models and, moreover, that a Poincar\'e-covariant treatment of the structure of each, independent system is crucial to its sound explanation, interpretation, and understanding.
It is worth continuing this elucidation by canvassing properties of the $(3/2,3/2^+)$ $\Delta(1232)$ and its first $(3/2,3/2^+)$ excitation, the $\Delta(1600)$.

A detailed analysis of $\Delta(1232)3/2^+$ and $\Delta(1600) 3/2^+$ Poincar\'e covariant wave functions, obtained in the $q(qq)$ framework, is presented in Ref.\,\cite{Liu:2022ndb}.  In these cases, the diquark structure is straightforward.
Namely, since it is impossible to construct an isospin $3/2$ system using an isoscalar diquark, such systems can only contain axialvector and/or vector diquarks.
Furthermore, since vector diquarks are (\emph{a}) effectively more massive than axialvector diquarks and (\emph{b}) have opposite parity to the baryon under consideration, then only axialvector diquarks need be considered.
These points are elucidated, \emph{e.g}., in Refs.\,\cite{Eichmann:2016jqx, Chen:2019fzn}.

\begin{figure}[t]
\begin{minipage}[t]{\textwidth}
\leftline{%
\includegraphics[clip, width=0.5\textwidth]{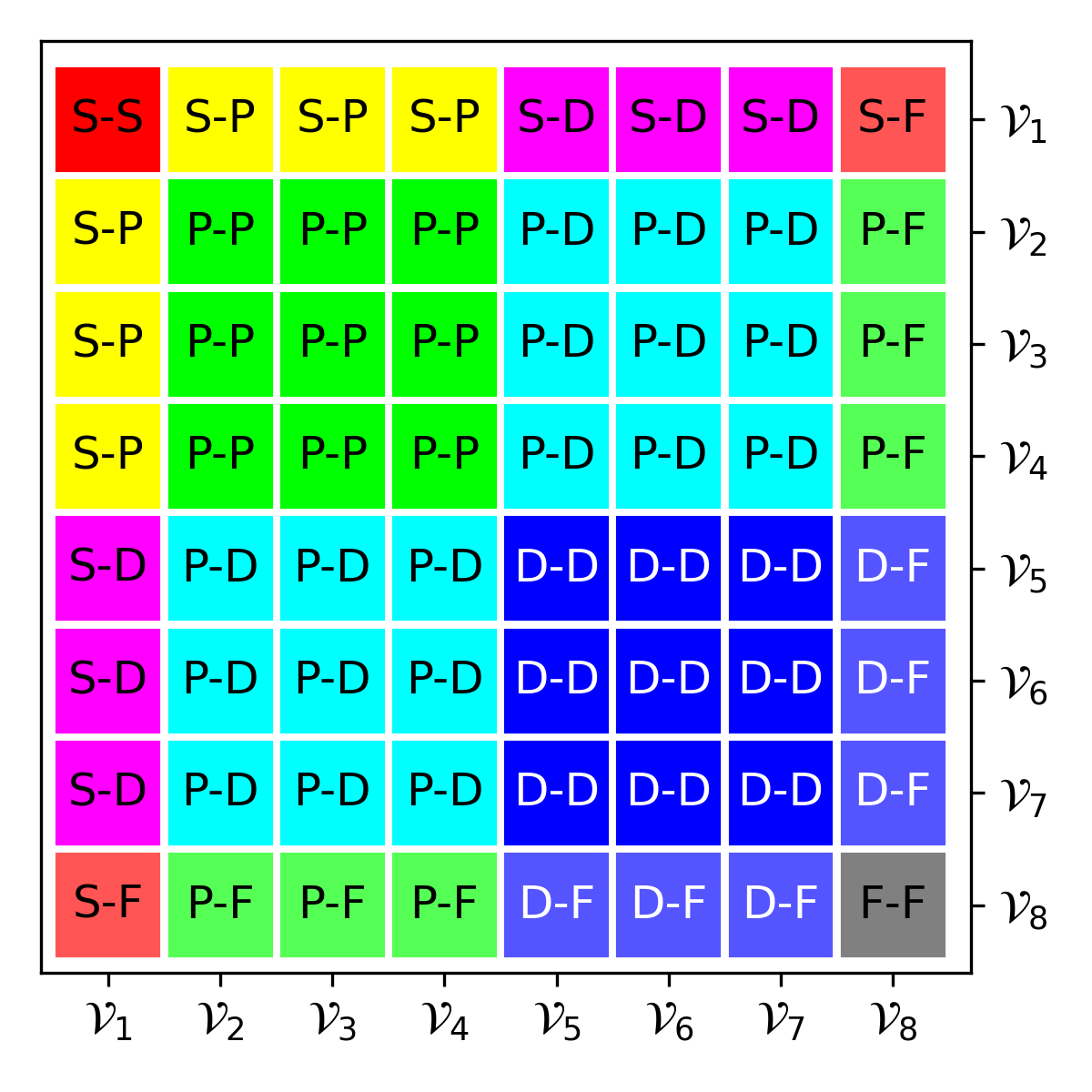}}
\end{minipage}\vspace*{-17em}
\rightline{
\begin{minipage}[t]{0.45\textwidth}
\caption{\label{F5}
Legend for interpretation of $J=3/2$ baryon rest-frame quark + diquark angular momentum decompositions, which also identifies interference between the distinct orbital angular momentum basis components.
These states can possess ${\mathsf S}$, ${\mathsf P}$, ${\mathsf D}$, and ${\mathsf F}$-wave components.
The axes labels refer to distinct components of the $q(qq)$ wave function, which are explicitly detailed in
Ref.\,\cite[Eq.\,(15)]{Liu:2022ndb}.
}\vspace*{2em}
\end{minipage}
}
\end{figure}

In continuing, it is first worth noting that the $\Delta(1600) 3/2^+$ wave function, like that of the Roper, exhibits features that justify its interpretation as the first radial excitation of the $\Delta(1232) 3/2^+$.  This interpretation is supported, \emph{e.g}., by the appearance of a single zero in the zeroth Chebyshev moment of each component of that wave function which has significant magnitude \cite{Liu:2022ndb}.
On the other hand, as we will see, and again akin to the $N(1440) 1/2^+$, the $\Delta(1600)3/2^+$ is not simply a radial excitation.

\begin{figure}[t]
\hspace*{-1ex}\begin{tabular}{lcl}
\large{\textsf{A}} & & \large{\textsf{B}}\\[-0.7ex]
%
\includegraphics[clip, width=0.45\textwidth]{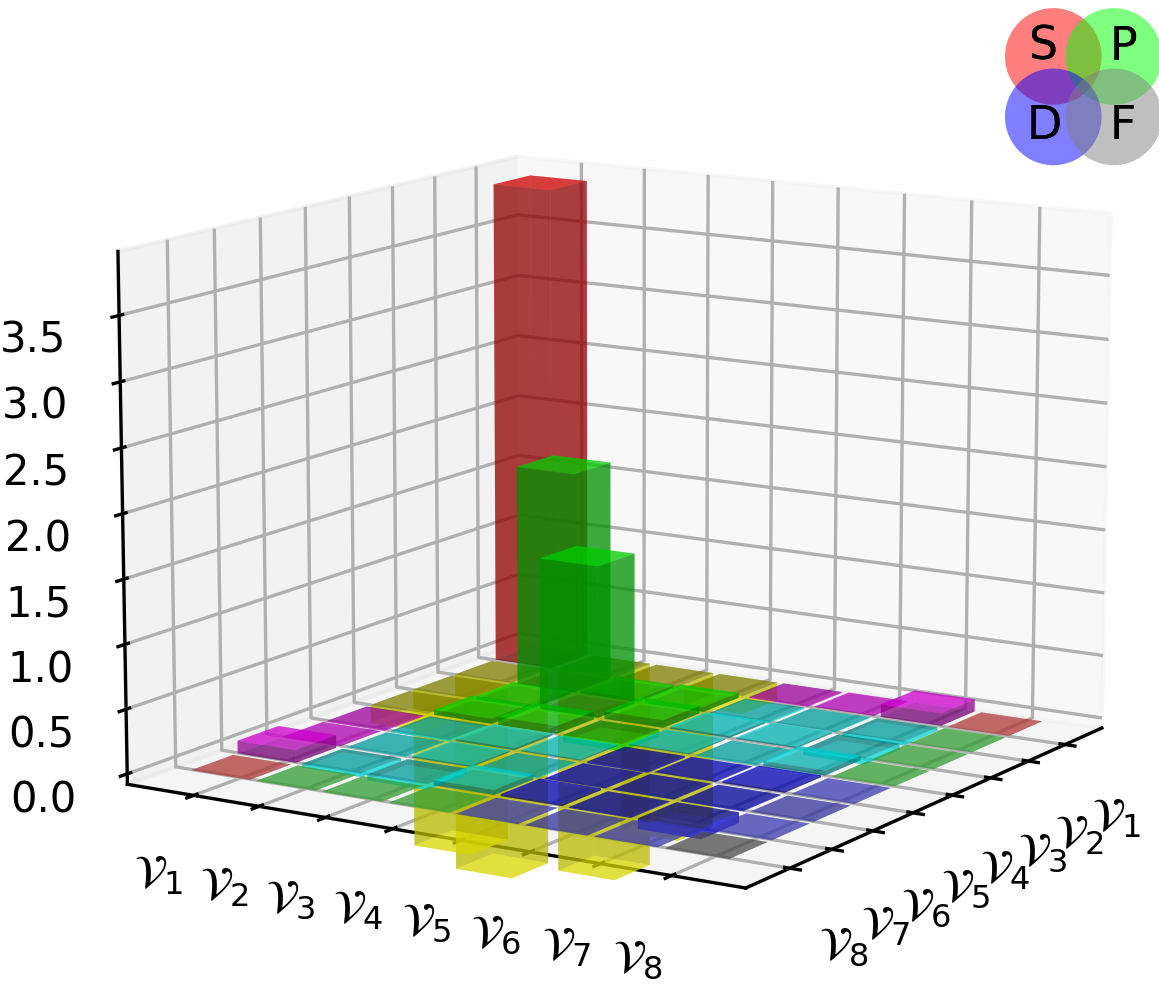} & \hspace*{1em} &
\includegraphics[clip, width=0.45\textwidth]{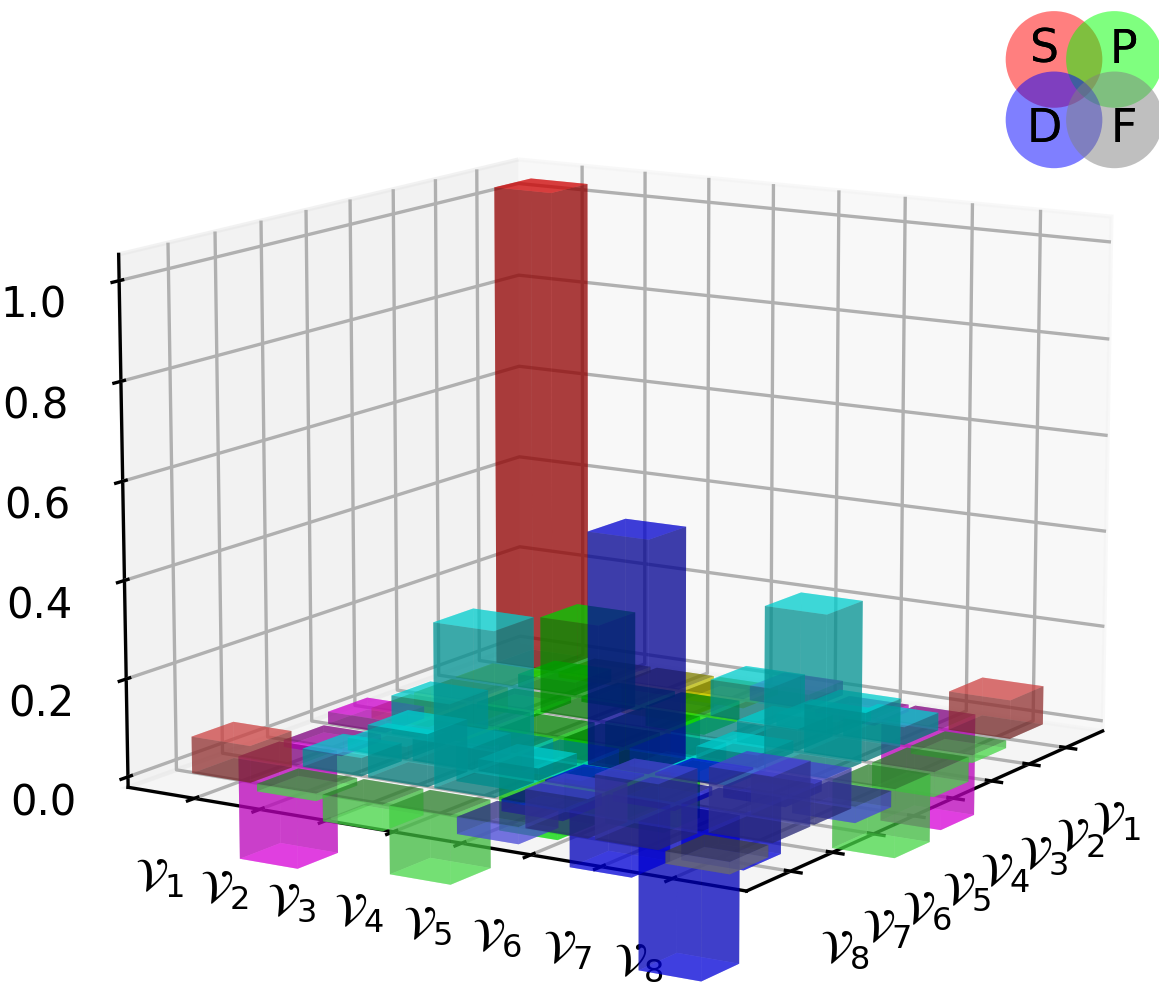}
\end{tabular}
\caption{\label{F6}
Contributions of the various quark + diquark orbital angular momentum components to the canonical normalisation of the Poincar\'e-covariant wave function of a $J=3/2$ baryon after rest-frame projection: there are both positive (above plane) and negative (below plane) contributions to the overall positive normalisation.
{\sf Panel A}.  $\Delta(1232)$.
{\sf Panel B}.  First $J=3/2^+$ excitation of the $\Delta(1232)$, which can be identified with the $\Delta(1600)$ resonance.
%
%
}
\end{figure}

Rest-frame $q(qq)$ angular momentum decompositions are available for $\Delta(1232) 3/2^+$, $\Delta(1600) 3/2^+$ \cite{Liu:2022ndb}.  Using the assignments displayed in Fig.\,\ref{F5}, the structures of these wave functions are displayed in Fig.\,\ref{F6}.
Considering Fig.\,\ref{F6}\,A, one sees that, evaluated in the rest frame, the canonical normalisation of the $\Delta(1232) 3/2^+$ is largely determined by $\mathsf S$-wave components, although there are significant, constructive $\mathsf P$ wave contributions and also strong $\mathsf S\otimes \mathsf P$-wave destructive interference terms.  The structural picture of the $\Delta(1232) 3/2^+$ communicated by this image has been confirmed by comparisons with data on the $\gamma^\ast+p \to \Delta(1232) 3/2^+$ transition form factors \cite{Eichmann:2011aa, Segovia:2014aza, Lu:2019bjs}.  (An update of these predictions will soon be available from the $q(qq)$ framework in Ref\,\cite{Cheng:2025yij}.)

Turning to Fig.\,\ref{F6}\,B, it is evident that, although $\mathsf S$-wave contributions are dominant in the $\Delta(1600) 3/2^+$, there are prominent $\mathsf D$-wave components, significant $\mathsf P \otimes \mathsf D$-wave interference contributions, and numerous $\mathsf F$-wave induced interference terms.
(Amplified higher partial waves are also seen in related $3$-body studies of the $\Delta(1600) 3/2^+$ \cite{Eichmann:2016hgl, Qin:2018dqp}.)
Here it is important to stress that this $q(qq)$ structural picture of the $\Delta(1600) 3/2^+$ has been used to predict $\gamma^\ast + p \to \Delta(1600)$ transition form factors \cite{Lu:2019bjs}.
Those predictions were confirmed in an analysis of $\pi^+ \pi^- p$ electroproduction data collected at JLab; see Refs.\,\cite[Sec.\,4.2]{Achenbach:2025kfx}, \cite{Mokeev:2023zhq}.
It is worth noting, too, that the predictions in Ref.\,\cite{Lu:2019bjs} have been updated, using the framework in Ref.\,\cite{Cheng:2025yij}.
The new results will be reported elsewhere.  Again, they confirm those which are already available.

That confirmation leads us to reiterate a material point.
Namely, as with the $N(1440) 1/2^+$, some practitioners deny a connection between the $\Delta(1600) 3/2^+$ and the first $3/2^+$ excitation of the $\Delta(1232) 3/2^+$ \cite{Leinweber:2024psf}.
Instead, they suggest that it is some sort of molecular state following an analysis based on a framework that attributes the same structure to the $N(1440) 1/2^+$.
Such pictures do not incorporate the truly complex wave function displayed in Fig.\,\ref{F6}\,B, with its strong orbital angular momentum correlations.
They could only become viable if tested against data on the $\gamma^\ast + p \to \Delta(1600)$ transition.
Such a validation seems far away; hence, one should today remain wary of these molecular pictures; especially in the face of the $q(qq)$ structural predictions that deliver agreement with the electroproduction data \cite{Lu:2019bjs}.

\section{$\Delta(1700)3/2^-$}
The refined $q(qq)$ framework described and deployed in Ref.\,\cite{Cheng:2025yij} has already been employed in numerous applications.  For instance, predictions for light-front transverse charge and magnetisation densities for the proton and neutron and their dressed valence-quark constituents can be found in Ref.\,\cite{Xu:2025ups}.
Analyses of transition form factors will subsequently be released.  Herein, therefore, we will only provide a sketch of predictions for $\Delta(1700)3/2^-$ electroproduction form factors.
The calculational procedure does not differ from that used in connection with, \emph{e.g}., the $\Delta(1600) 3/2^+$ transition form factors \cite{Lu:2019bjs}.

\begin{figure}[t]
\hspace*{-1ex}\begin{tabular}{lcl}
\large{\textsf{A}} & & \large{\textsf{B}}\\[-0.7ex]
%
\includegraphics[clip, width=0.45\textwidth]{F6A.png} & \hspace*{1em} &
\includegraphics[clip, width=0.45\textwidth]{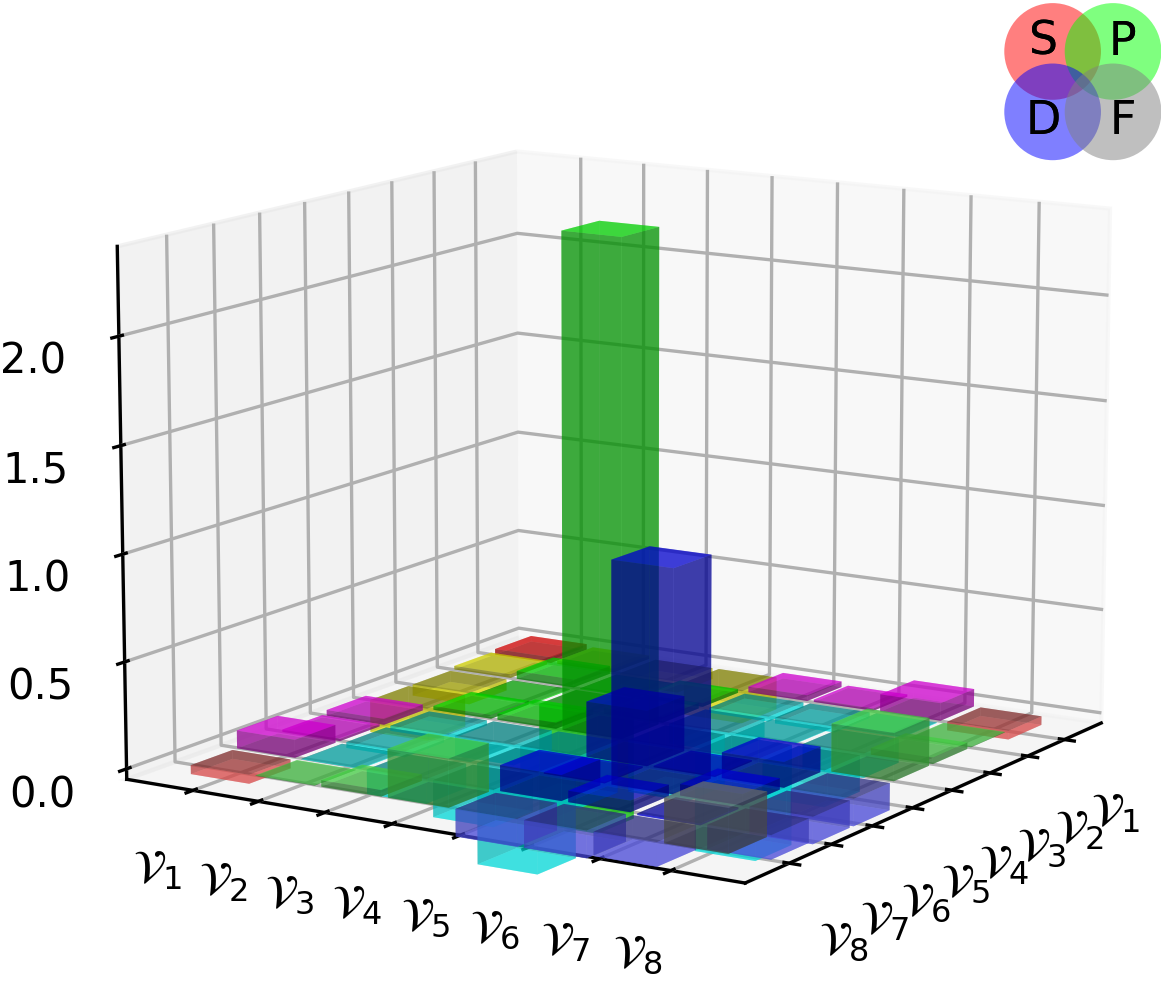}
\end{tabular}
\caption{\label{F7}
Contributions of the various quark + diquark orbital angular momentum components to the canonical normalisation of the Poincar\'e-covariant wave function of a $J=3/2$ baryon after rest-frame projection: there are both positive (above plane) and negative (below plane) contributions to the overall positive normalisation.
{\sf Panel A}.  $\Delta(1232)$.
{\sf Panel B}.  First $J=3/2^-$ excitation of the $\Delta(1232)$, which can be identified with the $\Delta(1700)3/2^-$ resonance.
%
%
The interpretation legend is provided in Fig.\,\ref{F5}.
}
\end{figure}

We begin by considering the Poincar\'e-covariant wave function of the $\Delta(1700)3/2^-$.
Possessing opposite parity to the $\Delta(1232) 3/2^+$ ground state, then quark potential models typically identify the $\Delta(1700)\tfrac{3}{2}^-$ as an $L=1$ orbital angular momentum excitation of the $\Delta(1232)$, so, ${\mathsf P}$-wave.  What does quantum field theory have to say?

The first point is that since this system does have opposite parity to the $\Delta(1232) 3/2^+$ ground state, it can contain negative-parity vector diquarks.  Notwithstanding that, they play a very minor role; regarding Ref.\,\cite[Table~II]{Liu:2022ndb}, one sees that they contribute just 5\% of the amplitude and their omission leaves the mass practically unchanged.  Thus, like the ground state, the $\Delta(1700)3/2^-$ is built, almost entirely, from isovector-axialvector diquarks.

Considering now the pointwise behaviour, one finds that most of the zeroth Chebyshev moment projections of the $\Delta(1700) 3/2^-$ wave function possess a zero.
When a zero exists, it lies within the domain $\tfrac{1}{3}{\rm fm} - \tfrac{1}{2}{\rm fm}$, \emph{i.e}., at length-scales smaller than the bound-state radius.  This is similarly true of $(\tfrac{1}{2},\tfrac{1}{2}^\pm)$ bound-state wave functions \cite[Figs.\,4, 5]{Chen:2017pse} and also vector mesons \cite[Fig.\,5]{Qin:2011xq}.
These things indicate that the $\Delta(1700) 3/2^-$ is much more than merely an $L=1$ excitation of the $\Delta(1232) 3/2^+$.

Such an observation leads one to ask whether the $\Delta(1700)\tfrac{3}{2}^-$ has \emph{any} of the characteristics of an orbital angular momentum excitation.
This is addressed by considering the rest-frame $q(qq)$ angular momentum decomposition of the $\Delta(1700) 3/2^-$ wave function, which is displayed in Fig.\,\ref{F7} and compared directly with that of the $\Delta(1232) 3/2^+$.
Working with this image, one may consider how the mass obtained by solving the Faddeev equation develops according to the inclusion or exclusion of different partial waves; see Ref.\,\cite[Table~III]{Liu:2022ndb}.  Concerning the $\Delta(1700) 3/2^-$, the lightest mass is obtained by keeping only $\mathsf P$ waves.  The net effect of adding the other components is a 4\% increase in the mass.  One must therefore expect that the state is dominated by the $\mathsf P$ wave components.

This expectation is confirmed by Fig.\,\ref{F7}\,B, which shows that the dominant contribution to the
$ \Delta(1700) 3/2^-$ canonical normalisation is provided by $\mathsf P$-wave components.
In addition, there are $\mathsf D$-wave pieces, $\mathsf P \otimes \mathsf D$ interference is evident, and also some $\mathsf D \otimes \mathsf F$ contributions.
Thus, in potentially accidental agreement with quark potential models, the $ \Delta(1700) 3/2^-$ is primarily a ${\mathsf P}$-wave state, which, nonetheless, also possesses measurable ${\mathsf S}$-, ${\mathsf D}$-wave components.

\begin{figure}[t]
\hspace*{-1ex}\begin{tabular}{lcl}
\large{\textsf{A}} & & \large{\textsf{B}}\\[-0.7ex]
%
\includegraphics[clip, width=0.48\textwidth]{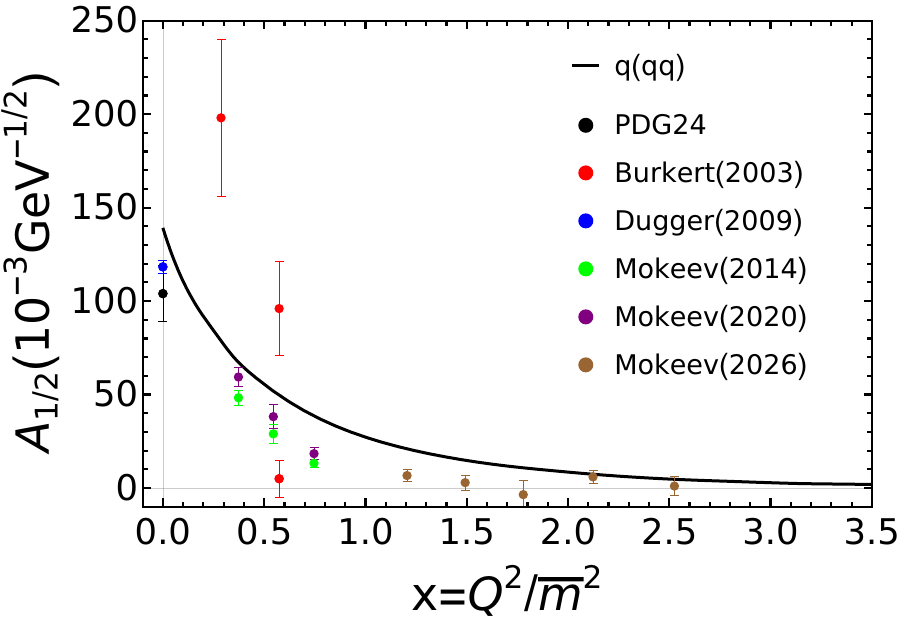} & \hspace*{-0.5em} &
\includegraphics[clip, width=0.48\textwidth]{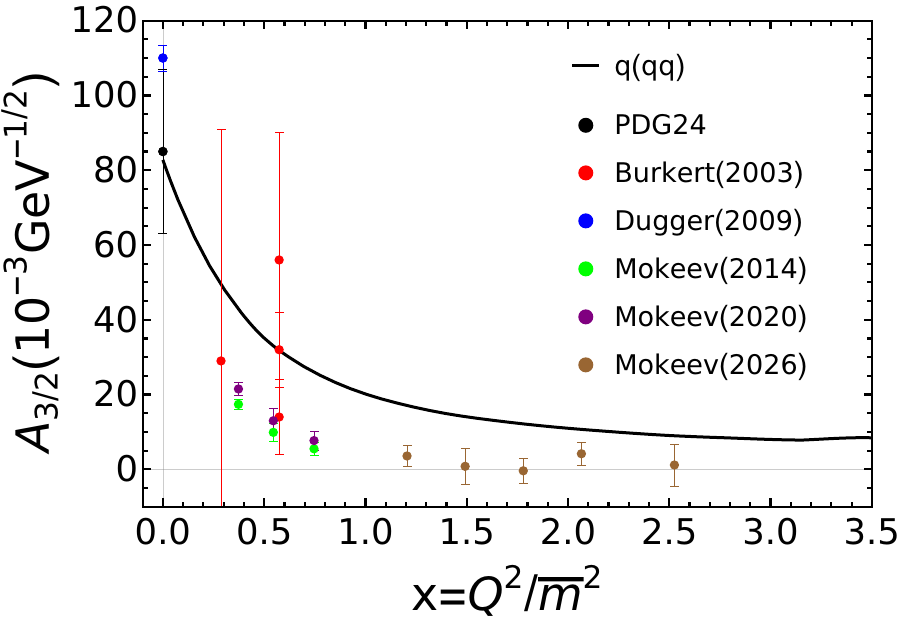}
\end{tabular}
\hspace*{-1ex}\begin{tabular}{lcl}
\large{\textsf{C}} & & $\,$\\[-0.7ex]
%
\includegraphics[clip, width=0.48\textwidth]{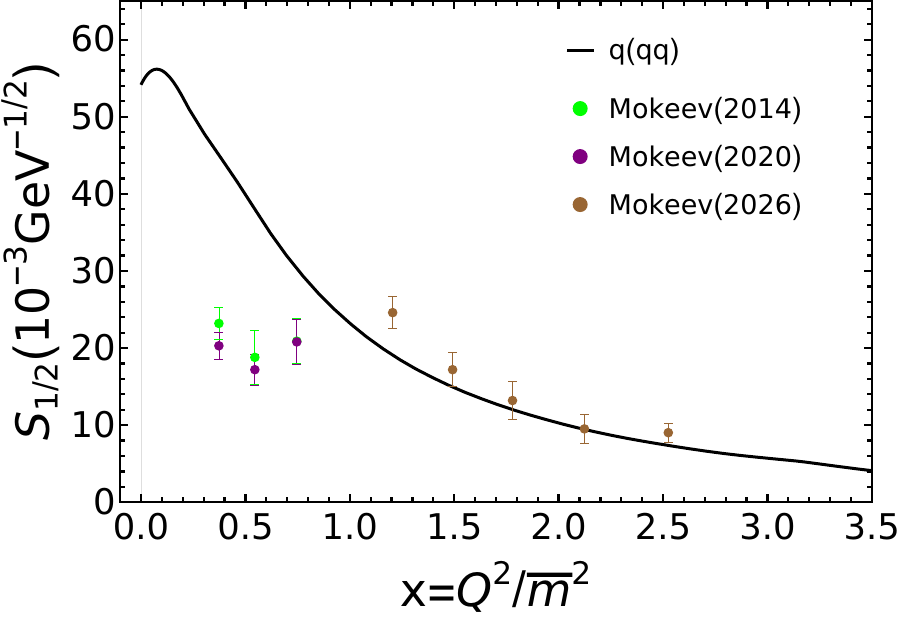} & \hspace*{-0.5em} &
\end{tabular}
\caption{\label{F8}
Helicity amplitudes: $\gamma^\ast p \to \Delta(1700) 3/2^-$ transition ($2 \bar m = m_N+m_{\Delta(1700)}$).
Solid black curve -- prediction obtained using the $q(qq)$ framework in Ref.\,\cite{Cheng:2025yij}.
Data -- Refs.\,\cite{Burkert:2002zz, CLAS:2009tyz, Mokeev:2013kka, Mokeev:2020hhu, Mokeev:2026}.
{\sf Panel A}.  $A_{1/2}$.
{\sf Panel B}.  $A_{3/2}$.
{\sf Panel C}.  $S_{1/2}$.
%
%
%
}
\end{figure}

One way to test the wave function in Fig.\,\ref{F7}\,B is to use it as the basis for predictions of the $\gamma^{\ast} N \to \Delta(1700)$ transition amplitudes and a subsequent comparison with relevant data.
In order to present such a comparison, we first note that a $\gamma^{\ast} N \to \Delta$ transition is described by three Poincar\'e-invariant form factors \cite{Jones:1972ky}: magnetic-dipole,
$G_{M}^{\ast}$;
electric quadrupole, $G_{E}^{\ast}$;
and Coulomb (longitudinal)
quadrupole, $G_{C}^{\ast}$.
They arise through consideration of the $N\to \Delta$
transition current:
\begin{equation}
J_{\mu\lambda}(K,Q) =
\Lambda_{+}(P_{f})R_{\lambda\alpha}(P_{f})i\gamma_{5}\Gamma_{\alpha\mu}(K,Q)\Lambda_
{+}(P_{i}),
\label{eq:JTransition}
\end{equation}
where: $P_{i}$, $P_{f}$ are, respectively, the incoming nucleon and outgoing $\Delta$
momenta, with $P_{i}^{2}=-m_{N}^{2}$, $P_{f}^{2}=-m_{\Delta}^{2}$; the incoming
photon momentum is $Q=(P_{f}-P_{i})$;
$K=(P_{i}+P_{f})/2$; and
$\Lambda_{+}(P_{i})$, $\Lambda_{+}(P_{f})$ are, respectively, positive-energy
projection operators for the nucleon and $\Delta$, with the Rarita-Schwinger tensor
projector $R_{\lambda\alpha}(P_f)$ arising in the latter connection.

In order to succinctly express $\Gamma_{\alpha\mu}(K,Q)$, we define
\begin{equation}
\check K_{\mu}^{\perp} = {\cal T}_{\mu\nu}^{Q} \check{K}_{\nu}
= (\delta_{\mu\nu} - \check{Q}_{\mu} \check{Q}_{\nu}) \check{K}_{\nu},
\end{equation}
with $\check{K}^{2} = 1 = \check{Q}^{2}$, in which case
\begin{align}
& \Gamma_{\alpha\mu}(K,Q)  \nonumber \\
& =
\mathpzc{k}
\left[\frac{\lambda_m}{2\lambda_{+}}(G_{M}^{\ast}-G_{E}^{\ast})\gamma_{5}
\varepsilon_{\alpha\mu\gamma\delta} \check K^\perp_{\gamma}\check{Q}_{\delta}  -
G_{E}^{\ast} {\cal T}_{\alpha\gamma}^{Q} {\cal T}_{\gamma\mu}^{\check K^\perp}
- \frac{i\varsigma}{\lambda_m}G_{C}^{\ast}\check{Q}_{\alpha} 
\check K^\perp_{\mu}
\right],
\label{eq:Gamma2Transition}
\end{align}
where
$\mathpzc{k} = \sqrt{(3/2)}(1+m_\Delta/m_N)$,
$\varsigma = Q^{2}/[2\Sigma_{\Delta N}]$,
$\lambda_\pm = \varsigma + t_\pm/[2 \Sigma_{\Delta N}]$
with $t_\pm = (m_\Delta \pm m_N)^2$,
$\lambda_m = \sqrt{\lambda_+ \lambda_-}$,
$\Sigma_{\Delta N} = m_\Delta^2 + m_N^2$, $\Delta_{\Delta N} = m_\Delta^2 - m_N^2$.

In terms of the Poincar\'e-invariant form factors, the directly measured $J^P = 3/2^-$ helicity amplitudes are expressed as follows:
\begin{subequations}
\begin{align}
A_{1/2} (Q^2) & = - \frac{1}{4 F_{1-}} \left[G_E^\ast(Q^2) - 3 G_M^\ast (Q^2)\right],\, \\
S_{1/2} (Q^2) & = - \frac{1}{\sqrt{2} F_{1-}} \frac{|{\bf q}|}{2 m_\Delta} G_C^\ast(Q^2),\, \\
A_{3/2} (Q^2) & = -\frac{\sqrt{3}}{4 F_{1-}} \left[G_E^\ast (Q^2) + G_M^\ast(Q^2)  \right]\,,
\label{half3_minus}
\end{align}
\end{subequations}
where
\begin{equation}
F_{1 -} = \frac{m_N}{|{\bf q}|} \frac{2 m_N}{(m_\Delta - m_N)}
\sqrt{\frac{\mathpzc M}{4 \pi \alpha}} \sqrt{\frac{Q^2_-}{4 m_N m_\Delta}}\,,
\label{eqFl1a}
\end{equation}
with $\alpha$ the quantum electrodynamics fine structure constant,
$\mathpzc M = [m_\Delta^2 - m_N^2]/[2 m_\Delta]$,
and the magnitude of the photon three-momentum  is
$|{\bf q}|= \sqrt{Q_+^2 Q_-^2}/[2 m_\Delta]$,  $Q_\pm^2 = (m_\Delta \pm m_N)^2 + Q^2$.

$\Delta(1700) 3/2^-$ transition electrocoupling data are available from JLab \cite{Burkert:2002zz, CLAS:2009tyz, Mokeev:2013kka, Mokeev:2020hhu, Mokeev:2026}.
So, in Fig.\,\ref{F8}, those data are compared with predictions for these electrocoupling helicity amplitudes that were obtained using the $q(qq)$ framework in Ref.\,\cite{Cheng:2025yij}.  Evidently, and uniformly on $x\gtrsim 2$, \emph{i.e}., above the domain on which meson-baryon final state interactions may play a role, the predictions agree well with data, as was the case for the Roper resonance; see Sec.\,\ref{SecRoper}.  This agreement confirms the viability of the $ \Delta(1700) 3/2^-$ wave function in Fig.\,\ref{F8}\,B.

\begin{figure}[t]
\begin{minipage}[t]{\textwidth}
\leftline{%
\includegraphics[clip, width=0.45\textwidth]{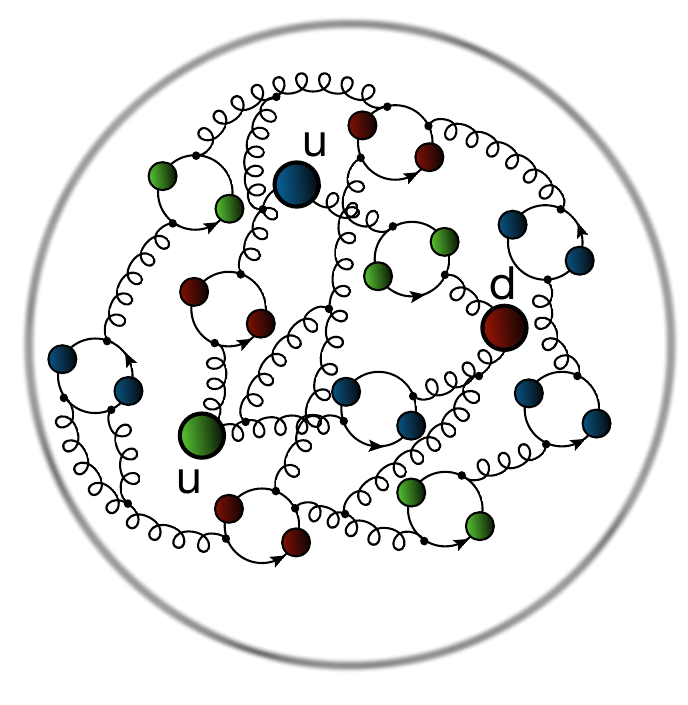}}
\end{minipage}\vspace*{-15.5em}
\rightline{
\begin{minipage}[t]{0.45\textwidth}
\caption{\label{F9}
Light-quark baryons contain three light valence quarks, and infinitely many gluon and sea-quark partons, drawn here as ``springs'' and closed loops, respectively.  This image is a sketch of the proton: two valence $u$ quark partons and one valence $d$ quark parton.  Adding strangeness, the number of possible baryon states increases by a factor of at least three.}\vspace*{2em}
\end{minipage}
}
\end{figure}

\section{Summary and Perspective}
The first baryon resonance, $\Delta(1232) 3/2^+$, was discovered in the early 1950s, \emph{i.e}., roughly eighty years ago.  Ten years later, the Roper resonance joined the collection.  Today, the number of known baryon resonances is very large; see, \emph{e.g}., Ref.\,\cite[Fig.\,15.4]{ParticleDataGroup:2024cfk}, which is, notably. not exhaustive.  As baryons, these states are the must fundamental three-body systems in Nature -- they must all be understood, not only the isolated ground state nucleon.

On the other hand, supposing both that quantum chromodynamics (QCD) is the theory underlying strong interactions and the degrees of freedom evident in the defining Lagrangian are all that one knows, then these states are not three-body systems at all.  Instead, they are complicated infinitely-many body systems built from gluon and quark partons, whose only three-body feature is their valence quark content; see Fig.\,\ref{F9}.  Numerical simulations of lattice-regularised QCD approach the baryon problem from this direction: ground states are readily accessible \cite{Durr:2008zz}, but resonances are not \cite{Briceno:2017max}.

Herein, we have primarily discussed an alternative to lattice simulations, namely, the continuum Schwinger function approach to hadron bound states.  In this case, baryons \emph{are} three-body systems, in the sense that they are built from three dressed-quark quasiparticle degrees-of-freedom, whose characteristics are an expression of emergent hadron mass (EHM).  EHM is a powerful concept, whose beginning is found in the dynamical generation of a gluon mass scale through gluon self-interactions; see Ref.\,\cite{Cornwall:1981zr} and citations thereof.  Exploiting EHM via QCD's Dyson-Schwinger equations (DSEs) -- the quantum Euler-Lagrange equations for the theory --  all glue and sea partons are sublimated into the dressed valence quarks at the hadron scale \cite{Yao:2025fnb}.  Consequently, they are described by a momentum dependent mass function, $M(k^2)$, which is large at infrared momenta, \emph{viz}.\ for light quarks, $M(0) \approx m_N/3$; see Ref.\,\cite[Sec.\,2C]{Roberts:2021nhw}.

\begin{figure}[t]
\centerline{%
\includegraphics[clip, width=0.70\textwidth]{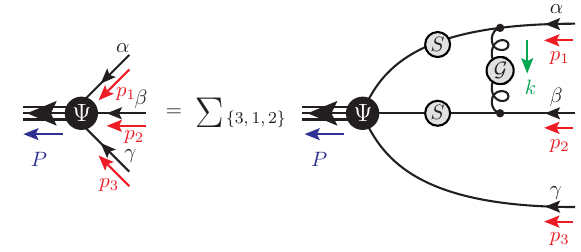}}
\caption{\label{F10}
Three-body Faddeev equation for a baryon with total momentum $P=p_1+p_2+p_3$, drawn in rainbow-ladder (RL) truncation, which is leading order in the systematic, symmetry-preserving DSE approximation scheme introduced in Refs.\,\cite{Munczek:1994zz, Bender:1996bb}.
Filled circle: Faddeev amplitude, $\Psi$, the matrix-valued solution, which involves 128 independent scalar functions.
Spring: dressed-gluon interaction that mediates quark+quark scattering; see Ref.\,\cite{Qin:2011dd, Binosi:2014aea}.
Solid line: dressed-quark propagator, $S$, calculated from the rainbow gap equation.
Lines not adorned with a shaded circle are amputated.
Isospin symmetry is assumed.
The sum runs over each of the cases involving quark ``$i=1,2,3$'' as a spectator to the gluon exchange interaction.}
\end{figure}

The valence-quark quasiparticle picture has passed many experimental and theoretical tests, \emph{e.g}., delivering sound predictions for nucleon elastic electromagnetic and gravitational form factors  \cite{Cheng:2025yij, Yao:2024ixu} and a large array of unpolarised and polarised proton parton distribution and fragmentation functions \cite{Lu:2022cjx, Cheng:2023kmt, Yu:2024ovn, Xing:2025eip}.
The majority of studies have hitherto employed the quark + fully-interacting diquark, $q(qq)$, simplification of the baryon problem.  Herein, we have sketched some of its applications to the study of baryon resonance electroexcitation.
(Spectrum studies -- $q(qq)$ and $3$-body -- are provided, \emph{e.g}., in Refs.\,\cite{Eichmann:2016yit, Qin:2019hgk, Yin:2019bxe, Yin:2021uom}.)
As noted, too, other $q(qq)$ predictions for electroexcitation amplitudes will soon be released.

The near future will see completion of work on the study of resonance excitation using Poincar\'e-covariant wave functions obtained from the direct $3$-body Faddeev equation sketched in Fig.\,\ref{F10}.  Where comparisons are already available, $q(qq)$ and $3$-body predictions are compatible.
Notwithstanding that, the approximations implicit in the $q(qq)$ simplification should be checked in all areas.
Then the tightest possible links can be forged with QCD and science can arrive at a robust understanding of the nucleon and all its resonance, \emph{viz}.\ the spectrum and structure of Nature's most fundamental three-body systems.

\begin{CJK*}{UTF8}{gbsn}
\section*{Acknowledgment}
We are grateful to I.\ I.\ Strakovsky for inviting us to contribute to this volume; and to
C.\  Chen (陈晨)
and
Z.-Q.\ Yao (姚照千) for constructive comments.
%
Work supported by:
National Natural Science Foundation of China (grant nos.\ 12135007, 12205149);
%
%
and
Natural Science Foundation of Anhui Pro\-vince, grant no.\ 2408085QA028.
\end{CJK*}


\end{document}